\documentclass[aps,prd,superscriptaddress,showpacs,preprint,amsmath,amssymb]{revtex4}
\usepackage{graphicx, bm}
\usepackage{float}
\usepackage[usenames]{color}

\usepackage{subcaption}
\begin{document}

\draft

\title{Sensitivities on the anomalous quartic $\gamma\gamma\gamma\gamma$ and $\gamma \gamma\gamma Z$ couplings at the CLIC }

\author{E. Gurkanli\footnote{egurkanli@sinop.edu.tr}}
\affiliation{\small Department of Physics, Sinop University, Turkey.\\}

\date{\today}

\begin{abstract}

It is essential to directly investigate the self-couplings of gauge bosons in the Standard Model (SM) due to its non-Abelian nature, as these couplings play a significant role in comprehending the gauge structure of the model. The discrepancies between the Standard Model's expectations and the measured value of gauge boson self-couplings serve as strong evidence for new physics phenomena extending beyond the Standard Model. Such deviations provide valuable insights into the nature of new physics and potentially lead to a deeper understanding of fundamental particles and their interactions. This study examines the sensitivities of anomalous couplings associated with dimension-8 operators that affect the $\gamma\gamma \gamma\gamma$ and $Z\gamma\gamma\gamma$ quartic vertices. The study focuses on the process $e^- \gamma \to e^-\gamma\gamma$ with the incoming photon under Weizsäcker–Williams approximation at the stage-3 scenario of Compact Linear Collider (CLIC) that refers to a CoM energy of 3 TeV. Due to the CLIC options, we take into account both unpolarized and 
$\mp80\%$ polarized electron beam with the related integrated luminosities of ${\cal L}=5, 4, 1$ $\rm ab^{-1}$ under the systematic uncertainties of $\delta_{sys}=0, 3, 5$. Obtained sensitivities on the anomalous quartic gauge couplings (aQGCs) for the process $e^- \gamma \to e^-\gamma\gamma$ at $\sqrt{s}= 3$ TeV and various polarizations are improved by a factor of 2-200 times better for the couplings $f_{T,j}/\Lambda^4$ compared with the experimental results reported by CMS Collaboration in Ref.\cite{JHEP10-2021}

-----

\end{abstract}


\vspace{5mm}

\maketitle


\section{Introduction}

The non-Abelian nature of the SM implies the gauge boson's self-interactions as triple and quartic gauge couplings(TGC and QGC). Also, possible deviations from triple and quartic gauge interactions as anomalous TGC and QGC (aTGC and aQGC) have an essential role in checking the validity of SM and predicting the new physics contribution coming from the Beyond Standard Model(BSM)\cite{arXiv:2106.11082,PRD93-2016,JHEP10-2017,JHEP10-2021}. In the SM, neutral gauge boson $ZZ\gamma\gamma, Z\gamma\gamma\gamma, \gamma\gamma\gamma\gamma$ couplings are excluded. In this manner, the contributions from these vertices are sources of new physics and a sign beyond the Standard Model.

These effects can be determined with Effective Field Theory (EFT) by adding high-dimension operators to the SM Lagrangian. It provides a model-independent framework to parametrize the effects of new physics systematically. It is discussed in the papers \cite{CPC44-2020, PRD107-2023} to generate the neutral TGC (NTGC) $ZV\gamma$ $(V=\gamma,Z)$ by dim-8 operators. Numerous experimental and phenomenological studies have been conducted to explore high-dimensional gauge operators, both at present and future colliders, focusing on proton-proton, electron-proton, and electron-positron collisions \cite{Eboli-PRD93-2016, Marantis-JPCS2105-2021,Degrande-AP335-2013,Hao-PRD104-2021,Degrande-JHEP02-2014,Hays-JHEP02-2019,Gounaris-PRD65-2002,Wrishik-JHEP08-2022,Murphy-JHEP10-2020,Ellis-China64-2021,Murphy-JHEP04-2021,Green-RMP89-2017,PLB760-2016,JHEP12-2018,PLB540-2002,PRD70-2004,PRD62-2000,PRD88-2013,PRL113-2014,PRL115-2015,PRD96-2017,EPJC77-2017,PRD95-2017,PRD90-2014,PLB774-2017,PLB770-2017,JHEP06-2017,PRL120-2018,PLB795-2019,PLB798-2019,PLB812-2021,JHEP06-2020,PLB809-2020,PLB811-2020,Stirling,PRD89-2014,EPJC13-2000,PLB515-2001,Eboli,PRD75-2007,AHEP2016-2016,JHEP10-121-2021,EPJC81-2021,EPJP130-2015,PRD104-2021,JHEP06-142-2017,JPG26-2000,Eboli2,PRD81-2010,PRD85-2012,arXiv:0908.2020,PRD86-2012,Eboli4,arXiv:2109.12572,Han-PLB1998,Daniele-RNC1997,Boos-PRD1998,Boos-PRD2000,Beyer-EPJC2006,Christian-EPJC2017,Stephen-arxiv:9505252,Stirling-PLB1999,Belanger-PLB1992,Senol-AHEP2017,Barger-PRD1995,Cuypers-PLB1995,Cuypers-IJMPA1996,Ji-arXiv:2204.08195,ATLAS1,ATLAS2,CMS1,JPG2022,JPG2023,Koksal-son}. One of the current phenomenological studies aims to investigate anomalous $\gamma\gamma\gamma\gamma$ couplings via the $\mu^{+}\mu^{-} \to \mu^{+} \gamma \gamma \mu^{-}$ process at future muon collider with the center-of-mass energies $\sqrt{s}=$3, 14 and 100 TeV \cite{arXiv:2306.03653}. In this context, performing this work in CLIC will be a complementary study of future lepton colliders.

The current CLIC scenarios supposed 1.0 $\rm ab^{-1}$, 2.5 $\rm ab^{-1}$, and 5.0 $\rm ab^{-1}$ of luminosities with the center-of-mass energies of $\sqrt{s}=0.38$ TeV, 1.5 TeV, and 3 TeV, respectively \cite{franc}. Also, CLIC enables $\pm80\%$ polarisation options for the electron beam, but there is no option for positron polarisation. Here, unpolarized and polarized electron beam of $P_{e^-}=-80\%$, $80\%$ are pertinent with the luminosities of ${\cal L}=5$ ab$^{-1}$,  ${\cal L}=4$ ab$^{-1}$ and ${\cal L}=1$ ab$^{-1}$, respectively \cite{CLIC-1812.06018}. Polarization options provide a better signal-background ratio to reach better sensitivities on aQGC with high luminosities. Besides, the collisions at the fundamental level in the lepton collider provide a cleaner environment than the hadron colliders. This gives us an advantage of minor systematic uncertainties, easy analysis, and precise measurements.

The cross-section of any process consists of the cross-sections related to electron and positron beam polarizations ($P_{e^-}$,$P_{e^+}$) given in the following equation \cite{kek2017}.

\begin{eqnarray}
\label{}
\sigma\left(P_{e^+},P_{e^-}\right)=\frac{1}{4}\{\left(1+P_{e^+}\right)\left(1+P_{e^-}\right)\sigma_{RR}+\left(1-P_{e^+}\right)\left(1-P_{e^-}\right)\sigma_{LL}
\\ \nonumber
+\left(1-P_{e^+}\right)\left(1+P_{e^-}\right)\sigma_{RL}+\left(1+P_{e^+}\right)\left(1-P_{e^-}\right)\sigma_{LR}\}\,,
\end{eqnarray}

Here, $\sigma_{LL}$ and $\sigma_{RR}$ are the cross-sections of both left and right-handed electron-positron beams. Similarly, $\sigma_{RL}$ and $\sigma_{LR}$ represent the cross-conditions of the beam polarizations.

In this study, the process $e^- \gamma \to e^-\gamma\gamma$ with a CoM energy $\sqrt{s}=3$ TeV at CLIC for both unpolarized and polarized $P_{e^-}=-80\%$, $80\%$ is taken into account to examine the new physics effects via aQGC under various systematic uncertainties. The paper consists of the following parts. Section II defines the dim-8 EFT for the anomalous $Z\gamma\gamma\gamma$ and $\gamma\gamma\gamma\gamma$ couplings. Applied analyzing techniques on the process and obtained sensitivities on the $f_{T,j}/\Lambda^4$ couplings at $95\%$ Confidence Level (C.L.) are discussed in Section-III and Section-IV, respectively. Finally, we sum up our results in Section V.

\section{Dim-8 operators for the anomalous $Z\gamma\gamma\gamma$ and $\gamma\gamma\gamma\gamma$ couplings}

EFT in the SM is a framework that extends the reach of the SM of particle physics. It provides a systematic way to describe the effects of new physics(NP). Here, the starting point is to identify the effective operators. The EFT Lagrangian consists of the SM Lagrangian plus an infinite tower of higher-dimension operators, which are suppressed by a cut-off scale. The EFT Lagrangian is given as follows.

\begin{equation}
{\cal L}_{eff}={\cal L}_{SM}+\sum_{i}\frac{c_i^{(6)}}{\Lambda^2}{\cal O}_i^{(6)}
+\sum_{j}\frac{c_j^{(8)}}{\Lambda^4}{\cal O}_j^{(8)}+...,
\end{equation}

The processes, including aQGCs, can be produced via tri-boson productions and the vector boson scattering (VBS) processes. Here, VBS processes refer to the scattering of two vector bosons with each other that can produce quartic gauge boson interactions. On the other hand, Tri-boson production refers to the production of three vector bosons. Compared with the tri-boson processes, the VBS processes exhibit a higher sensitivity to aQGCs.\cite{ATLAS1,ATLAS2,CMS1}.

In this study, we focused on the dim-8 anomalous $Z\gamma\gamma\gamma$ and $\gamma\gamma\gamma\gamma$ quartic vertices using Effective Lagrangian techniques by adding dim-8 operators to the generic SM Lagrangian. Generally, the formalism for aQGC has been discussed in the literature  \cite{Eboli1,Eboli,Degrande,Eboli3,Eboli-PRD101-2020}. The effective Lagrangian, including the dim-8 operators for quartic couplings, is as follows.

\begin{equation}
{\cal L}_{eff}= {\cal L}_{SM} +\sum_{k=0}^2\frac{f_{S, k}}{\Lambda^4}O_{S, k} +\sum_{i=0,1,2,3,4,5,7}^{7}\frac{f_{M, i}}{\Lambda^4}O_{M, i}+\sum_{j=0,1,2,5,6,7,8,9}^{}\frac{f_{T, j}}{\Lambda^4}O_{T, j},
\end{equation}

Here, $O_{S, k}$, $O_{M, i}$ and $O_{T, j}$ is the operators of dim-8 and $\frac{f_{S, k}}{\Lambda^4}$, $\frac{f_{M, i}}{\Lambda^4}$ and $\frac{f_{T, j}}{\Lambda^4}$ are the related parameters running by the effective operators. In Eq.3, there are three types of aQGC operators.

The first set of operators ($\frac{f_{S, k}}{\Lambda^4} O_{S, k}$) is contribute with the covariant derivative which induces the $ZZZZ$, $WWZZ$ and $WWWW$ couplings. On the other hand, $\frac{f_{T, j}}{\Lambda^4}$ and $\frac{f_{M, i}}{\Lambda^4}$  operators are contributing with Gauge boson field strength tensors and both field (Mixed field), respectively. Below, the aQGC operators corresponding to these three type operators are presented \cite{Degrande}. \\

$\bullet$ Scalar field:

\begin{eqnarray}
O_{S, 0}&=&[(D_\rho\Phi)^\dagger (D_\sigma\Phi)]\times [(D^\rho\Phi)^\dagger (D^\sigma\Phi)],  \\
O_{S, 1}&=&[(D_\rho\Phi)^\dagger (D^\rho\Phi)]\times [(D_\sigma\Phi)^\dagger (D^\sigma\Phi)],  \\
O_{S, 2}&=&[(D_\rho\Phi)^\dagger (D_\sigma\Phi)]\times [(D^\sigma\Phi)^\dagger (D^\rho\Phi)].
\end{eqnarray}

$\bullet$  Tensor field:

\begin{eqnarray}
O_{T, 0}&=&Tr[\widehat{W}_{\sigma\lambda} \widehat{W}^{\sigma\lambda}]\times Tr[\widehat{W}_{\alpha\beta}\widehat{W}^{\alpha\beta}],  \\
O_{T, 1}&=&Tr[\widehat{W}_{\lambda\mu} \widehat{W}^{\nu\beta}]\times Tr[\widehat{W}_{\nu\beta}\widehat{W}^{\lambda\mu}],  \\
O_{T, 2}&=&Tr[\widehat{W}_{\lambda\nu} \widehat{W}^{\nu\sigma}]\times Tr[\widehat{W}_{\sigma\mu}\widehat{W}^{\mu\lambda}],  \\
O_{T, 5}&=&Tr[\widehat{W}_{\lambda\sigma} \widehat{W}^{\lambda\sigma}]\times B_{\mu\nu}B^{\mu\nu},  \\
O_{T, 6}&=&Tr[\widehat{W}_{\lambda\mu} \widehat{W}^{\nu\sigma}]\times B_{\nu\sigma}B^{\lambda\mu},  \\
O_{T, 7}&=&Tr[\widehat{W}_{\lambda\nu} \widehat{W}^{\nu\sigma}]\times B_{\sigma\mu}B^{\mu\lambda},  \\
O_{T, 8}&=&B_{\sigma\lambda} B^{\sigma\lambda}B_{\mu\nu}B^{\mu\nu},  \\
O_{T, 9}&=&B_{\lambda\nu} B^{\nu\sigma}B_{\sigma\mu}B^{\mu\lambda}.
\end{eqnarray}

$\bullet$ Mixed field:

\begin{eqnarray}
O_{M, 0}&=&Tr[\widehat{W}_{\nu\lambda} \widehat{W}^{\nu\lambda}]\times [(D_\sigma\Phi)^\dagger (D^\sigma\Phi)],  \\
O_{M, 1}&=&Tr[\widehat{W}_{\nu\lambda} \widehat{W}^{\lambda\sigma}]\times [(D_\sigma\Phi)^\dagger (D^\nu\Phi)],  \\
O_{M, 2}&=&[B_{\nu\lambda} B^{\nu\lambda}]\times [(D_\sigma\Phi)^\dagger (D^\sigma\Phi)],  \\
O_{M, 3}&=&[B_{\nu\lambda} B^{\lambda\sigma}]\times [(D_\sigma\Phi)^\dagger (D^\nu\Phi)],  \\
O_{M, 4}&=&[(D_\nu\Phi)^\dagger \widehat{W}_{\sigma\lambda} (D^\nu\Phi)]\times B^{\sigma\lambda},  \\
O_{M, 5}&=&[(D_\nu\Phi)^\dagger \widehat{W}_{\sigma\lambda} (D^\lambda\Phi)]\times B^{\sigma\nu} + h.c. ,  \\
O_{M, 7}&=&[(D_\nu\Phi)^\dagger \widehat{W}_{\sigma\lambda} \widehat{W}^{\sigma\nu} (D^\lambda\Phi)].
\end{eqnarray}

In above equations, the subscripts $S$, $M$, $T$ refer to scalar (or longitudinal), $M$ mixed field and $T$ transversal. In these equations, $\Phi$ is the Higgs doublet, $D_\mu\Phi=(\partial_\mu + igW^j_\mu \frac{\sigma^j}{2}+ \frac{i}{2}g'B_\mu )\Phi$ is the covariant derivatives of the Higgs field and finally Pauli matrices are denoted by $\sigma^j (j=1,2,3)$. Here, $B_{\mu\nu}$ and $\widehat{W}_{\mu\nu} \equiv \sum_{i}^{}\frac{1}{2} \sigma^i W^{i} $ represent the gauge field strength tensors.

As mentioned above, we aimed at the dim-8 aQGC operators. A list of quartic vertices altered with the dim-8 operators is given in Table I. In this study, the evaluated process is sensitive to T-type operators. Because of this, we only consider the $f_{T,j}/\Lambda^4$ parameters with $j=0,1,2,5,6,7,8,9$. Obtained sensitivities on $f_{T,8}/\Lambda^4$ and $f_{T,9}/\Lambda^4$ parameters are important due to the relation on the electroweak neutral bosons only. On the other hand, the analytical expressions of the anomalous $Z \gamma\gamma\gamma$ coupling and related anomalous parameters for the process $e^- \gamma \to e^-\gamma\gamma$ are given in Eqs. (22)-(25) \cite{PRD104-2021,Ji-arXiv:2204.08195}.

\begin{eqnarray}
V_{Z\gamma \gamma \gamma,1}=F^{\mu\nu}F_{\mu\nu}F^{\alpha\beta}Z_{\alpha\beta},
\end{eqnarray}
\begin{eqnarray}
V_{Z\gamma \gamma \gamma,2}=F^{\mu\nu}F_{\nu\alpha}F^{\alpha\beta}Z_{\beta\mu}.
\end{eqnarray}

Here, $Z^{\mu\nu}=\partial^\mu Z^{\nu}-\partial^\nu Z^{\mu}$. The related coefficients are given in the following equations.

\begin{eqnarray}
\alpha_{Z\gamma\gamma\gamma,1}=\frac{c_{W}^3 s_{W}}{\Lambda^4}( f_{T,5}+ f_{T,6}-4f_{T,8})+\frac{c_{W} s_{W}^3}{\Lambda^4}(f_{T,0}+f_{T,1}-f_{T,5}-f_{T,6}),
\end{eqnarray}

\begin{eqnarray}
\alpha_{Z\gamma\gamma\gamma,2}=\frac{c_{W}^3 s_{W}}{\Lambda^4}( f_{T,7}- 4f_{T,9})+\frac{c_{W} s_{W}^3}{\Lambda^4}(f_{T,2}-f_{T,7}).
\end{eqnarray}

\begin{table}{H}
\caption{The aQGCs related with dim-8 operators are represented with $\bullet$ \cite{Degrande}.}
\begin{center}
\begin{tabular}{|l|c|c|c|c|c|c|c|c|c|}
\hline
& $WWWW$ & $WWZZ$ & $ZZZZ$ & $WW\gamma Z$ & $WW\gamma \gamma$ & $ZZZ\gamma$ & $ZZ\gamma \gamma$ & $Z \gamma\gamma\gamma$ & $\gamma\gamma\gamma\gamma$ \\
\hline
\cline{1-10}
$O_{S0}$, $O_{S1}$                     & $\bullet$ & $\bullet$ & $\bullet$ &   &   &   &   &   &    \\
$O_{M0}$, $O_{M1}$, $O_{M7}$ & $\bullet$ & $\bullet$ & $\bullet$ & $\bullet$ & $\bullet$ & $\bullet$ & $\bullet$ &   &    \\
$O_{M2}$, $O_{M3}$, $O_{M4}$, $O_{M5}$ &   & $\bullet$ & $\bullet$ & $\bullet$ & $\bullet$ & $\bullet$ & $\bullet$ &   &    \\
$O_{T0}$, $O_{T1}$, $O_{T2}$           & $\bullet$ & $\bullet$ & $\bullet$ & $\bullet$ & $\bullet$ & $\bullet$ & $\bullet$ & $\bullet$ & $\bullet$  \\
$O_{T5}$, $O_{T6}$, $O_{T7}$           &   & $\bullet$ & $\bullet$ & $\bullet$ & $\bullet$ & $\bullet$ & $\bullet$ & $\bullet$ & $\bullet$  \\
$O_{T8}$, $O_{T9}$                     &   &   & $\bullet$ &   &   & $\bullet$ & $\bullet$ & $\bullet$ & $\bullet$  \\
\hline
\end{tabular}
\end{center}
\end{table}

\section{Analysis for the aQGC via the process $e^- \gamma \to e^-\gamma\gamma$ }

Linear colliders offer a powerful means to explore physics beyond the Standard Model by utilizing $\gamma\gamma$ and $e^{-}\gamma$ interactions. In these colliders, emitted photons from incoming electrons undergo minimal scattering angles with the beam pipe, resulting in almost real, low-virtuality photons. The Weizsacker-Williams approximation, also known as the Equivalent Photon Approximation (EPA), plays a pivotal role in phenomenological studies, allowing researchers to estimate cross sections for processes like $e^{-}\gamma \to X$ by studying the primary $e^{-}e^{+}\to e^{-}\gamma e^{+}\to Xe^{+}$ interaction. This "X" represents particles observed in the final state, and these interactions benefit from exceptionally clean experimental conditions due to the collision of elementary particles (not composed of smaller constituents), free from hadronic activities. With the emission of an elastic photon, incoming charged particles, be they electrons or protons, scatter at small angles and evade detection by central detectors. This creates a distinctive missing energy signature known as a forward large-rapidity gap in the forward region of the central detector. Using forward detectors with well-synchronized central detectors is highly effective in mitigating background. It's worth noting that the program at CLIC has embraced forward physics and incorporated extra detectors located at varying distances from the interaction point, enhancing their capabilities\cite{CLIC}. In the context of this study, the EPA, commonly referred to as the Weizsacker-Williams Approximation (WWA)\cite{WWA1,WWA2}, serves as the primary theoretical tool for estimating sensitivity in the total cross-section of the process $e^-\gamma \to e^{-} \gamma \gamma$ to analyze the potential of CLIC-based electron-photon colliders to probe the anomalous $Z\gamma\gamma\gamma$ and $\gamma\gamma\gamma\gamma$ quartic vertices in the stage-3 scenario with electron polarizations of $P_{e^-}=-80\%$, $0\%$, $80\%$. 

Here, the incoming photon from the positron beam is taken under Weizsäcker–Williams approximation (WWA). The following equation presents the photon's spectrum emitted by a positron beam. \cite{Budnev,Chen2}.

\begin{eqnarray}
f_{\gamma}(x)=\frac{\alpha}{\pi E_{e}}\{[\frac{1-x+x^{2}/2}{x}]log(\frac{Q_{max}^{2}}{Q_{min}^{2}})-\frac{m_{e}^{2}x}{Q_{min}^{2}}
\\ \nonumber (1-\frac{Q_{min}^{2}}{Q_{max}^{2}})-\frac{1}{x}[1-\frac{x}{2}]^{2}log(\frac{x^{2}E_{e}^{2}+Q_{max}^{2}}{x^{2}E_{e}^{2}+Q_{min}^{2}})\} \nonumber \\
\end{eqnarray}

Here $x=E_{\gamma}/E_{e}$ and $Q_{max}^{2}$ is the photon's maximum virtuality. On the other hand, the  expression of $Q_{min}^{2}$ is given as:

\begin{eqnarray}
Q_{min}^{2}=\frac{m_{e}^{2}x^{2}}{1-x}.
\end{eqnarray}

Using the methodology, the cross-section of the $e^- \gamma \to e^-\gamma\gamma$ process
at CLIC can be calculated as follows.

\begin{eqnarray}
\sigma=\int f_{\gamma}(x) d\hat{\sigma}dE_{1}.
\end{eqnarray}

Possible Feynman diagrams of the process $e^- \gamma \to e^-\gamma\gamma$ are shown in Fig.1. Here, the black dots show the effective vertices. In the analysis, we aimed at the cross-sections to constrain the anomalous $\frac{f_{T, j}}{\Lambda^4}$ parameters related to $Z\gamma\gamma\gamma$ and $\gamma\gamma\gamma\gamma$ quartic vertices. The total cross-section of the process $e^- \gamma \to e^-\gamma\gamma$, including anomalous interaction, is composed as the summation of SM, purely new physics contribution, and the interference part. This is given as follows.

\begin{eqnarray}
\sigma_{Tot}\Biggl( \sqrt{s}, \frac{f_{T,j}}{\Lambda^{4}}\Biggr)&=& \sigma_{SM}( \sqrt{s} )
+\sigma_{(INT)}\Biggl( \sqrt{s}, \frac{f_{T,j}}{\Lambda^{4}}\Biggr)   \nonumber\\
&+& \sigma_{NP}\Biggl(\sqrt{s}, \frac{f^2_{T,j}}{\Lambda^{8}} \Biggr), \nonumber\\
 \hspace{3mm} j= 0-2, 5-9,    \nonumber\\
\end{eqnarray}

\noindent In the above equation, $\sigma_{SM}$ is the SM cross-section. On the other hand, $\sigma_{(INT)}$ consists of the contributions from the interference between SM and the EFT operators. Lastly, $\sigma_{NP}$ is the cross-section purely contributed by the EFT operators. During the analysis, every coefficient is taken to zero at a time.

To evaluate the total cross-section $\sigma_{Tot}\biggl(\sqrt{s},\frac{f_{T,j}}{\Lambda^{4}}\biggr)$ of the process $e^- \gamma \to e^-\gamma\gamma$, we used the MadGraph5\_aMC@NLO\cite{MadGraph}. The operators described in Equations (7)-(14) were embedded into MadGraph5\_aMC@NLO using the Feynrules package\cite{AAlloul}, which served as a Universal FeynRules Output (UFO) module\cite{CDegrande}.

In the first step, we evaluate the SM background and the signals cross-sections for each anomalous parameter at a time with the no-cut situation to determine the optimized kinematic cuts for the next step.  
In Fig.2, we give the transverse momentum of the final state photons $p_{T}^{\gamma}$. It must be mentioned that the given transverse momentum is the scalar sum of all photons. Here, $p_{T}^{\gamma}$ cut is very decisive, and a chosen value of 350 GeV is ideal for the separation area for the signals and SM background. The rest of the selected cuts came with the default values of the MadGraph5\_aMC@NLO. $p_{T}^{l}$ and $\eta^{\gamma}$ are the transverse momentum of the final state leptons and the pseudorapidities of the photons, respectively. We consider $p_{T}^{l} > 10$ GeV and  $\eta^{\gamma}< 2.5$ with labeled as Cut-1. Next, we applied an angular separations $(\Delta R =( (\Delta \phi)^2+ (\Delta \eta)^2)^{1/2}$ for the final state charged lepton and photons that are $\Delta R(\gamma, \gamma) > 0.4$, $\Delta R(\gamma, l) > 0.4$ which is labeled as Cut-2. The final state photons of the process $e^- \gamma \to e^-\gamma\gamma$ are useful to separate the signal and SM background events because, in the large values of $p_{T}^{\gamma}$, high dimensional operators can affect the transverse momentum of the photon. However, $p_{T}^{\gamma} > 350$ GeV is applied for the final state photons with labeled Cut-3.

\begin{table}{H}
\caption{Selected cuts for the process $e^- \gamma \to e^-\gamma\gamma$.}
\begin{tabular}{|c|c|c}
\hline
Kinematic cuts &       \\
\hline
\hline
Cut-1   & \, $p^l_T>10$ GeV , $|\eta^{\gamma}| < 2.5$    \\
\hline
Cut-2  & \multicolumn{1}{|c|}{$\Delta R(\gamma,\gamma) > 0.4$ , $\Delta R(\gamma,l) > 0.4$ }\\
\hline
Cut-3   & $p^\gamma_T > 350$ GeV\\
\hline
\end{tabular}
\end{table}

In Table III, a cut flow chart is given to see the effect of the selected kinematic cuts on the events for the signals and the SM background step-by-step. In table, we give the $f_{T,0}/\Lambda^4$ $f_{T,2}/\Lambda^4$, $f_{T,5}/\Lambda^4$, $f_{T,7}/\Lambda^4$, $f_{T,8}/\Lambda^4$ for signals and the SM background. All coupling values are taken 1 TeV$^{-4}$ one at a time. In the table, cut-0 refers to the default kinematic cuts to arrange the singularities and divergences in the phase space. After applying the selected cuts, we can see that the SM background events dramatically decrease compared with the signals. Also, we give the total cross-section as a function of anomalous parameters $f_{Tj}/\Lambda^4$ in Fig.3-5 to compare cross-sections of signals each other for the different polarizations of electron beams $P_{e^-}=-80\%, 0\%, 80\%$ after applying cuts given in Table-III.

Another critical topic is the systematic uncertainties while doing the analysis. We have obtained the sensitivities of anomalous parameters $f_{Tj}/\Lambda^4$ under the systematic uncertainties of $0\%$, $3\%$, and $5\%$ at CLIC. Possible sources of the systematic uncertainties are the jet-photon misidentification, integrated luminosities, and photon efficiencies. These are listed in detail in Table II, given in the reference \cite{PLB478-2000}. Other phenomenological studies also used the same systematic uncertainty values to obtain the sensitivities \cite{PLB478-2000,PRD98-2018,PRD98-015017-2018}. In the next section, we will take our analysis one step further and obtain the sensitivities on anomalous parameters of $\frac{f_{T, j}}{\Lambda^4}$.

\begin{table}{H}
\caption{Number of events for the process $e^- \gamma \to e^-\gamma\gamma$ for various couplings and SM background for every step of selected kinematic cuts given in Table II for the unpolarized electron beams.}
\begin{tabular}{|c|c|c|c|c|c|c|c}
\hline
\multicolumn{7}{|c|}{Unpolarized electron beams }\\
\hline
Kinematic Cuts &$f_{T0}/\Lambda^{4}$ = 1 TeV$^{-4}$& $f_{T2}/\Lambda^{4}$ & $f_{T5}/\Lambda^{4}$ & $f_{T7}/\Lambda^{4}$ &$f_{T8}/\Lambda^{4}$& Standard Model \\
\hline
\hline
Cut-0 &840 & 600 & 8100 & 2300 &293000& 530 \\
\hline
Cut-1  & 840 & 600 & 8100 & 2300 & 293000 & 530   \\
\hline
Cut-2  & 840 & 600 & 8100 & 2300 & 293000 & 530 \\
\hline
Cut-3  & 287 & 74 & 6799 & 1588 & 263424 & 12 \\
\hline
\end{tabular}
\end{table}

\section{Expected sensitivity on the anomalous $\frac{f_{T, j}}{\Lambda^4}$ parameters at the CLIC}

The chi-square method directly quantifies the parameters $f_ {T,j}/\Lambda^4$, $j=0-2, 5-9$ of new physics that is related to anomalous $Z\gamma\gamma\gamma$ and $\gamma\gamma\gamma\gamma$ couplings.

\begin{equation}
\chi^2(f_{T,j}/\Lambda^4)=\Biggl(\frac{\sigma_{SM}(\sqrt{s})-\sigma_{Total}(\sqrt{s}, f_{T,j}/\Lambda^4)}
{\sigma_{SM}(\sqrt{s})\sqrt{(\delta_{st})^2 + (\delta_{sys})^2}}\Biggr)^2,
\end{equation}

Here, $\sigma_{Total}(\sqrt{s} f_{T,j}/\Lambda^4)$ is the total cross section which is contributed by the anomalous couplings and the SM part. Besides, $\sigma_{SM}(\sqrt{s})$ is the SM cross-section. On the other hand, $\delta_{sys}$ and $\delta_{st}=\frac{1}{\sqrt{N_{SM}}}$ are the systematic and statistical error, respectively. In addition, $N_{SM}={\cal L}\times \sigma_{SM}$ is the number of events that ${\cal L}$ is the integrated luminosity.

\begin{table}{H}
\caption{Sensitivities at $95\%$ C.L. on the anomalous $Z\gamma\gamma\gamma$ and $\gamma\gamma\gamma\gamma$ couplings via the process $e^- \gamma \to e^-\gamma\gamma$ with the electron polarizations of $P_{e^-}=0\%, -80\%, 80\%$ under the systematic uncertainties of $\delta_{sys}=0\%, 3\%,
5\%$ are represented.}
\begin{tabular}{|c|c|c|c|c|}
\hline
$P_{e^-}$              &      & $0\%$       & $-80\%$    & $80\%$ \\
\hline
Couplings (TeV$^{-4}$) & & ${\cal L}=5$ ab$^{-1}$ & ${\cal L}=4$ ab$^{-1}$ & ${\cal L}=1$ ab$^{-1}$ \\
\hline
                      &$\delta=0\%$       &$[-0.56;0.62]\times10^{-1}$  &$[-4.92;4.48]\times10^{-2}$ &$[-1.93;1.90]\times10^{-1}$ \\
$f_{T0}/\Lambda^{4}$  &$\delta=3\%$       &$[-0.63;0.69]\times10^{-1}$  &$[-5.35;4.92]\times10^{-2}$ &$[-1.98;1.95]\times10^{-1}$ \\
                      &$\ \, \delta=5\%$ &$[-0.72;0.78]\times10^{-1}$  &$[-5.93;5.50]\times10^{-2}$ &$[-2.06;2.03]\times10^{-1}$ \\
\hline
                      &$\delta=0\%$       &$[-0.56;0.62]\times10^{-1}$  &$[-4.92;4.48]\times10^{-2}$ &$[-1.93;1.90]\times10^{-1}$ \\
$f_{T1}/\Lambda^{4}$  &$\delta=3\%$       &$[-0.63;0.69]\times10^{-1}$  &$[-5.35;4.92]\times10^{-2}$ &$[-1.98;1.95]\times10^{-1}$ \\
                      &$\ \, \delta=5\%$ &$[-0.72;0.78]\times10^{-1}$  &$[-5.93;5.50]\times10^{-2}$ &$[-2.06;2.03]\times10^{-1}$ \\
\hline
                      &$\delta=0\%$       &$[-0.98;1.56]\times10^{-1}$  &$[-0.94;1.03]\times10^{-1}$ &$[-3.99;3.92]\times10^{-1}$ \\
$f_{T2}/\Lambda^{4}$  &$\delta=3\%$       &$[-1.11;1.70]\times10^{-1}$  &$[-1.03;1.12]\times10^{-1}$ &$[-4.10;4.02] \times10^{-1}$ \\
                      &$\ \, \delta=5\%$ &$[-1.29;1.88]\times10^{-1}$  &$[-1.15;1.24]\times10^{-1}$ &$[-4.26;4.18]\times10^{-1}$ \\
\hline
                      &$\delta=0\%$       &$[-1.47;0.97]\times10^{-2}$  &$[-1.31;0.84]\times10^{-2}$ &$[-2.57;2.33]\times10^{-2}$ \\
$f_{T5}/\Lambda^{4}$  &$\delta=3\%$       &$[-1.60;1.10]\times10^{-2}$  &$[-1.19;1.11]\times10^{-2}$ &$[-2.63;2.29]\times10^{-2}$ \\
                      &$\ \, \delta=5\%$ &$[-1.77;1.27]\times10^{-2}$  &$[-1.32;1.24]\times10^{-2}$ &$[-2.74;2.50]\times10^{-2}$ \\
\hline
                     &$\delta=0\%$       &$[-1.47;0.97]\times10^{-2}$  &$[-1.31;0.84]\times10^{-2}$ &$[-2.57;2.33]\times10^{-2}$ \\
$f_{T6}/\Lambda^{4}$ &$\delta=3\%$       &$[-1.60;1.10]\times10^{-2}$  &$[-1.19;1.11]\times10^{-2}$ &$[-2.63;2.39]\times10^{-2}$ \\
                     &$\ \, \delta=5\%$ &$[-1.77;1.27]\times10^{-2}$  &$[-1.32;1.24]\times10^{-2}$ &$[-2.74;2.50]\times10^{-2}$ \\
\hline
                      &$\delta=0\%$       &$[-0.37;0.17]\times10^{-1}$  &$[-0.28;0.16]\times10^{-1}$ &$[-0.55;0.46]\times10^{-1}$ \\
$f_{T7}/\Lambda^{4}$  &$\delta=3\%$       &$[-0.40;0.20]\times10^{-1}$  &$[-0.30;0.19]\times10^{-1}$ &$[-0.56;0.48]\times10^{-1}$ \\
                      &$\ \, \delta=5\%$ &$[-0.44;0.23]\times10^{-1}$  &$[-0.33;0.21]\times10^{-1}$ &$[-0.58;0.50]\times10^{-1}$ \\
\hline
                      &$\delta=0\%$       &$[-1.79;2.05]\times10^{-3}$  &$[-2.88;2.49]\times10^{-3}$ &  $[-2.67;2.15]\times10^{-3}$\\
$f_{T8}/\Lambda^{4}$  &$\delta=3\%$       &$[-2.01;2.27]\times10^{-3}$  &$[-3.13;2.74]\times10^{-3}$ &  $[-2.74;2.21]\times10^{-3}$\\
                      &$\ \, \delta=5\%$ &$[-2.28;2.55]\times10^{-3}$  &$[-3.47;3.08]\times10^{-3}$ &  $[-2.84;2.31]\times10^{-3}$\\
\hline
                      &$\delta=0\%$       &$[-0.50;0.32]\times10^{-2}$  &$[-0.59;0.51]\times10^{-2}$ &$[-0.51;0.49]\times10^{-2}$ \\
$f_{T9}/\Lambda^{4}$  &$\delta=3\%$       &$[-0.54;0.36]\times10^{-2}$  &$[-0.65;0.56]\times10^{-2}$ &$[-0.53;0.50]\times10^{-2}$ \\
                      &$\ \, \delta=5\%$ &$[-0.60;0.42]\times10^{-2}$  &$[-0.72;0.63]\times10^{-2}$ &$[-0.55;0.52]\times10^{-2}$ \\
\hline
\end{tabular}
\end{table}

\begin{table}{H}
\caption{Sensitivities at $95\%$ C.L. on the anomalous $\gamma\gamma \gamma\gamma$ and $Z\gamma\gamma\gamma$ couplings via the process $e^- \gamma \to e^-\gamma\gamma$ for linear and linear+quadratic terms with the $P_{e^-}=0\%$ under the systematic uncertainties of $\delta_{sys}=0\%, 3\%,$ and $5\%$ are represented.}
\begin{tabular}{|c|c|c|c|c|}
\hline
\multicolumn{3}{|c|} {} & \multicolumn{1}{|c|} {Linear} & \multicolumn{1}{|c|} {Linear+Quadratic} \\
\hline
Couplings & Experimental Bound & Systematic Errors & Our Projection & Our Projection \\
\hline
& & $\delta=0\%$ & $[-1.11;1.11]\times10^{-1}$ & $[-0.56;0.62]\times10^{-1}$  \\

$f_{T0}/\Lambda^{4}$  & $[-5.70;5.46]$ & $\delta=3\%$ &$[-1.37;1.37]\times10^{-1}$  &$[-0.63;0.69]\times10^{-1}$ \\

& & $\delta=5\%$ & $[-1.75;1.75]\times10^{-1}$ &$[-0.72;0.78]\times10^{-1}$ \\
\hline
& & $\delta=0\%$ &$[-1.11;1.11]\times10^{-1}$ &$[-0.56;0.62]\times10^{-1}$ \\

$f_{T1}/\Lambda^{4}$  &$[-5.70;5.46]$ & $\delta=3\%$ & $[-1.37;1.37]\times10^{-1}$  & $[-0.63;0.69]\times10^{-1}$ \\

& & $\delta=5\%$ & $[-1.75;1.75]\times10^{-1}$ &$[-0.72;0.78]\times10^{-1}$  \\
\hline
& & $\delta=0\%$ &$[-2.58;2.58]\times10^{-1}$ & $[-0.98;1.56]\times10^{-1}$  \\

$f_{T2}/\Lambda^{4}$  &  $[-11.40;10.90]$ & $\delta=3\%$ &$[-3.20;3.20]\times10^{-1}$ & $[-1.11;1.70]\times10^{-1}$ \\

& &  $\delta=5\%$ & $[-4.09;4.09]\times10^{-1}$  & $[-1.29;1.88]\times10^{-1}$  \\
\hline
& &$\delta=0\%$ &$[-2.88;2.88]\times10^{-2}$ &$[-1.47;0.97]\times10^{-2}$  \\

$f_{T5}/\Lambda^{4}$  & $[-2.92;2.92]$ &  $\delta=3\%$ & $[-3.57;3.57]\times10^{-2}$ &$[-1.60;1.10]\times10^{-2}$ \\

& & $\delta=5\%$ &$[-4.54;4.54]\times10^{-2}$ & $[-1.77;1.27]\times10^{-2}$ \\
\hline
& &$\delta=0\%$ &$[-2.88;2.88]\times10^{-2}$ &$[-1.47;0.97]\times10^{-2}$  \\

$f_{T6}/\Lambda^{4}$  & $[-3.80;3.88]$ &  $\delta=3\%$ & $[-3.57;3.57]\times10^{-2}$ &$[-1.60;1.10]\times10^{-2}$ \\

& & $\delta=5\%$ &$[-4.54;4.54]\times10^{-2}$ & $[-1.77;1.27]\times10^{-2}$ \\
\hline
& &$\delta=0\%$ &$[-0.70;0.70]\times10^{-1}$ &$[-0.37;0.17]\times10^{-1}$  \\

$f_{T7}/\Lambda^{4}$  & $[-7.88;7.72]$ &  $\delta=3\%$ & $[-0.87;0.87]\times10^{-1}$ &$[-0.40;0.20]\times10^{-1}$  \\

& & $\delta=5\%$ &$[-1.11;1.11]\times10^{-1}$ & $[-0.44;0.23]\times10^{-1}$  \\
\hline
& &$\delta=0\%$ &$[-1.40;1.40]\times10^{-2}$ &$[-1.79;2.05]\times10^{-3}$  \\

$f_{T8}/\Lambda^{4}$  & $[-1.06;1.10]$ &  $\delta=3\%$ & $[-1.73;1.73]\times10^{-2}$ &$[-2.01;2.27]\times10^{-3}$  \\

& & $\delta=5\%$ &$[-2.21;2.21]\times10^{-2}$ & $[-2.28;2.55]\times10^{-3}$   \\
\hline
& &$\delta=0\%$ &$[-0.92;0.92]\times10^{-2}$ &$[-0.50;0.32]\times10^{-2}$\\

$f_{T9}/\Lambda^{4}$  & $[-1.82;1.82]$ &  $\delta=3\%$ & $[-1.13;1.13]\times10^{-2}$ &$[-0.54;0.36]\times10^{-2}$  \\

& & $\delta=5\%$ &$[-1.44;1.44]\times10^{-2}$ & $[-0.60;0.42]\times10^{-2}$  \\
\hline
\end{tabular}
\end{table}

 The anomalous parameters $f_{T,j}/\Lambda^4$ are obtained at the $95\%$ C.L. using the cross-sections of the process $e^- \gamma \to e^-\gamma\gamma$ after the selected cuts given in Table II for each coupling at a time. The process are performed in $\sqrt{s}=3$ TeV option with an integrated luminosities of ${\cal L}=1$ ${\rm ab^{-1}}$ ($P_{e^-}=80\%$), ${\cal L}=4$ ${\rm ab^{-1}}$ ($P_{e^-}=-80\%$) and ${\cal L}=5$ ${\rm ab^{-1}}$ ($P_{e^-}=0\%$) under the systematic uncertainties of $\delta_{sys}=0\%, 3\%, 5\%$ at CLIC collider.

Figs. 3-5 shows the variation of the anomalous couplings to the total cross-section after applying the selected cuts in Table II for different electron polarization options of $P_{e^-}=-80\%$, $P_{e^-}=0\%$ and $P_{e^-}=80\%$, respectively. In those figures, the dim-8 operators have strong dependencies and highly affect the cross-section to increase with its value.

Sensitivities on anomalous $Z\gamma\gamma\gamma$ and $\gamma\gamma\gamma\gamma$ couplings at $95\%$ C.L. with the process $e^- \gamma \to e^-\gamma\gamma$ at the CLIC are given in Table IV. The sensitivities are obtained for different electron polarization options $P_{e^-}=-80\%$, $0\%$, $80\%$  and under various systematic uncertainties of $\delta_{sys} = 0\%, 3\%, 5\%$. Here can be seen, $f_{T,5}/\Lambda^4$, $f_ {T,8}/\Lambda^4$, and $f_ {T,9}/\Lambda^4$ couplings have the restrictive sensitivities of  $[-1.31; 0.84] \times 10^{-2} \hspace{1mm} {\rm TeV^{-4}}$ , $ [-1.79; 2.05] \times 10^{-3} \hspace{1mm} {\rm TeV^{-4}}$ and $[-0.50; 0.32] \times 10^{-2} \hspace{1mm} {\rm TeV^{-4}}$, respectively.

In Figs. 6-13, we give the comparison of the anomalous $f_{T,j}/\Lambda^4$ parameters with the experimental results in Ref. \cite{JHEP10-2021} via the process $pp\to Z\gamma\gamma \to l^{+}l^{-}\gamma\gamma$ at $\sqrt{s}=13$ TeV and integrated luminosity of $\cal L =$ 137 fb$^{-1}$ that is reported by the CMS Collaboration. In those figures, we consider the electron polarizations of $P_{e^-}=80\%$ with the integrated luminosities of ${\cal L} = 0.1,0.5$ and $1$ $\rm ab^{-1}$. Similarly, $P_{e^-}=-80\%$ polarizations with the integrated luminosities of ${\cal L} = 0.1, 1, 4$ $\rm ab^{-1}$ and  ${\cal L} = 0.1, 1, 5$ $\rm ab^{-1}$ for unpolarized electron beam are also taken into account. Sensitivities on $f_{T,0}/\Lambda^4$, $f_{T,1}/\Lambda^4$, $f_{T,2}/\Lambda^4$, $f_{T,5}/\Lambda^4$, $f_{T,6}/\Lambda^4$ and $f_{T,7}/\Lambda^4$ couplings at $P_{e^-}=-80\%$ are more stringent than the obtained for the other options $P_{e^-}=0\%$ and $80\%$. On the other hand, $f_{T,8}/\Lambda^4$ and $f_{T,9}/\Lambda^4$ couplings have its best sensitivities at $P_{e^-}=0\%$. Also, in this study, we have extended our results with the sensitivities obtained via linear term (interference of dim8 operators with SM), given in Table V, and compared them with the full effects (linear+quadratic) and the experimental results. Obtained sensitivities via linear terms are worse than the full effects as expected. Here, the quadratic terms dominated the interference term. On the other hand, linear terms improved the experimental results by up to 200 times, as shown in Table V.

\section{CONCLUSIONS}

In this study, a specific class of interactions aQGCs are handled within the EFT framework, which is described by the non-Abelian gauge structure of the SM. In this context, the study may lead to an opportunity to test the validity of SM and give important clues to the presence of new physics. Additionally, lepton colliders are very suitable due to their compact design and cleaner background environment without any hadronic activity due to the nature of lepton collision compared with the LHC. These are the main motivations for performing this simulation in CLIC. Furthermore, the CLIC program gives high CoM energy up to a 3 TeV in the stage-3 scenario with high integrated luminosity. Also, we consider the electron polarization options of CLIC to see the effect on aQGC couplings.

With these motivations, we evaluate the process $e^- \gamma \to e^-\gamma\gamma$ at the CLIC to probe the dim-8 anomalous $Z\gamma\gamma\gamma$ and $\gamma \gamma\gamma\gamma$ couplings. Obtained sensitivities on dim-8 parameters $f_{T,j}/\Lambda^4$ that are between 2-200 times stronger than the experimental limits given in the Ref.\cite{JHEP10-2021}. On the other hand, anomalous $Z\gamma\gamma\gamma$ and $ZZ\gamma\gamma$ couplings are studied at CLIC via the process $e^{-}e^{+} \to Z\gamma\gamma$\cite{arXiv:2112.03948}. Obtained results in this study via the process $e^- \gamma \to e^-\gamma\gamma$ improved the sensitivities of aQGCs by a factor between 2-12 times compared with the Ref.\cite{arXiv:2112.03948}. Stringent bounds on the anomalous $f_{T,5,8,9}/\Lambda^4$ couplings are $\frac{f_{T5}}{\Lambda^{4}}= [-1.31; 0.84] \times 10^{-2} \hspace{1mm} {\rm TeV^{-4}}$ with ${\cal L}=4\hspace{0.8mm} \rm ab^{-1}$, $P_{e^-}=-80\%$ and $\delta_{sys}=0\%$, $\frac{f_{T8}}{\Lambda^{4}}= [-1.79; 2.05] \times 10^{-3} \hspace{1mm} {\rm TeV^{-4}}$ and $\frac{f_{T9}}{\Lambda^{4}}= [-0.50; 0.32] \times 10^{-2} \hspace{1mm} {\rm TeV^{-4}}$ with ${\cal L}=5\hspace{0.8mm} \rm ab^{-1}$, $P_{e^-}=0\%$ and $\delta_{sys}=0\%$. Consequently, $O_{T,8}$ and $O_{T,9}$ operators have the optimal sensitivities related on anomalous $Z\gamma\gamma\gamma$ and $\gamma \gamma\gamma\gamma$ couplings for the process $e^- \gamma \to e^-\gamma\gamma$. Also, we give the obtained sensitivities via linear terms and compare the full contributions (linear+quadratic) and the experimental results. Both linear and linear+quadratic contributions improved the experimental results remarkably.

\vspace{1.5cm}

\vspace{1cm}


\newpage

\begin{figure}[H]
\centerline{\scalebox{1.0}{\includegraphics{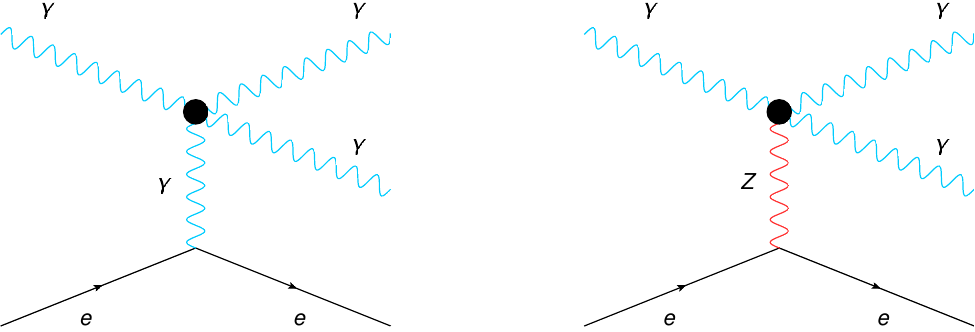}}}
\caption{ \label{fig:gamma} Diagrams for the process $e^- \gamma \to e^-\gamma\gamma$
involving the anomalous $Z\gamma\gamma \gamma$ and $\gamma\gamma\gamma\gamma$ couplings. New physics contributions are shown by a black circle.}
\end{figure}

\begin{figure}[H]
\centerline{\scalebox{0.9}{\includegraphics{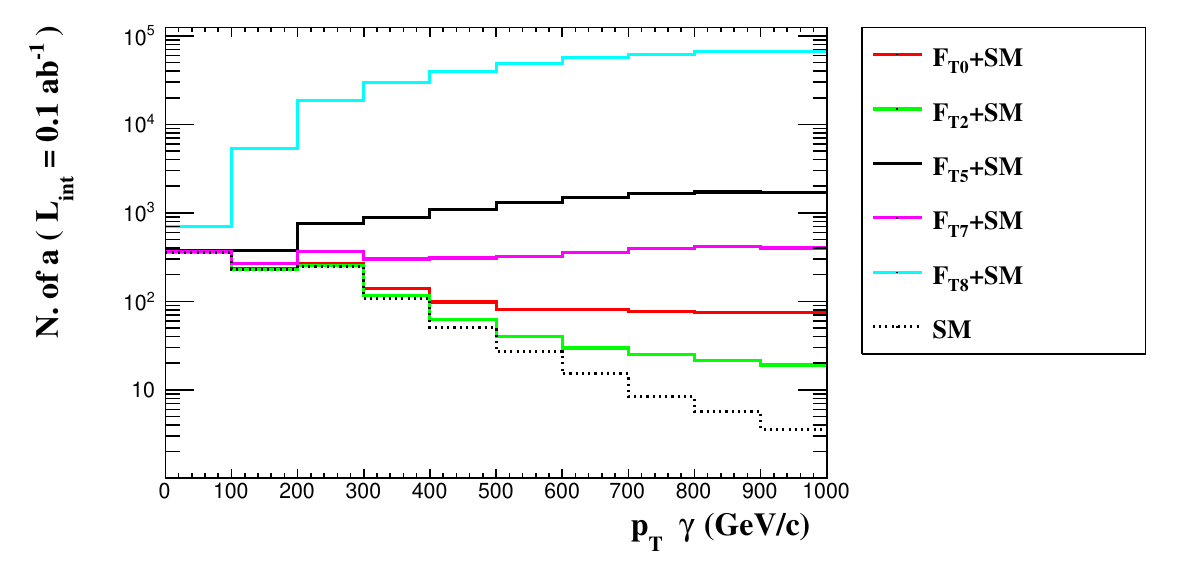}}}
\caption{ \label{fig:gamma}  The number of events as a function of the transverse momentum of the final state photon $p^{\gamma}_T$ for the process $e^- \gamma \to e^-\gamma\gamma$ and SM background at $\sqrt{s}=3$ TeV and $P_{e^-}=0\%$.}
\end{figure}

\begin{figure}[H]
\centerline{\scalebox{1.3}{\includegraphics{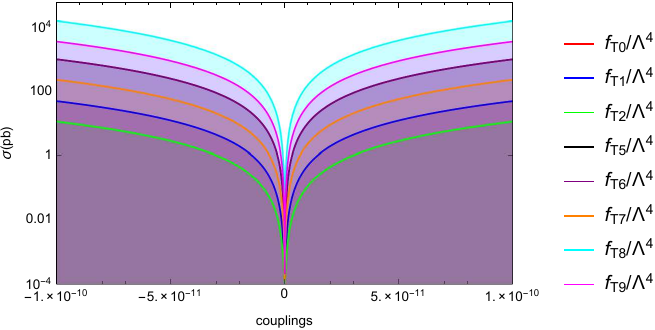}}}
\caption{ \label{fig:gamma} Cross-sections of the process $e^- \gamma \to e^-\gamma\gamma$
in terms of the anomalous parameters $f_ {T,j}/\Lambda^4$ for the polarized beams $P_{e^-}=-80\%$.}
\end{figure}

\begin{figure}[H]
\centerline{\scalebox{1.25}{\includegraphics{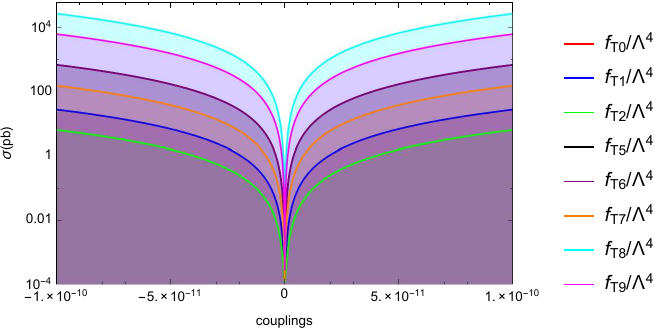}}}
\caption{ \label{fig:gamma} Same as in Fig. 3, but for $P_{e^-}=0\%$.}
\end{figure}

\begin{figure}[H]
\centerline{\scalebox{1.25}{\includegraphics{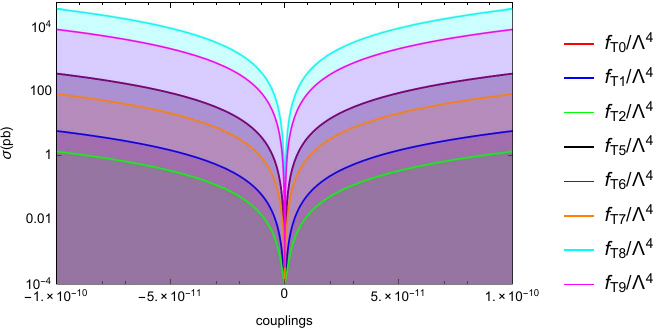}}}
\caption{ \label{fig:gamma} Same as in Fig. 3, but for $P_{e^-}=80\%$.}
\end{figure}

\begin{figure}[H]
\centerline{\scalebox{0.9}{\includegraphics{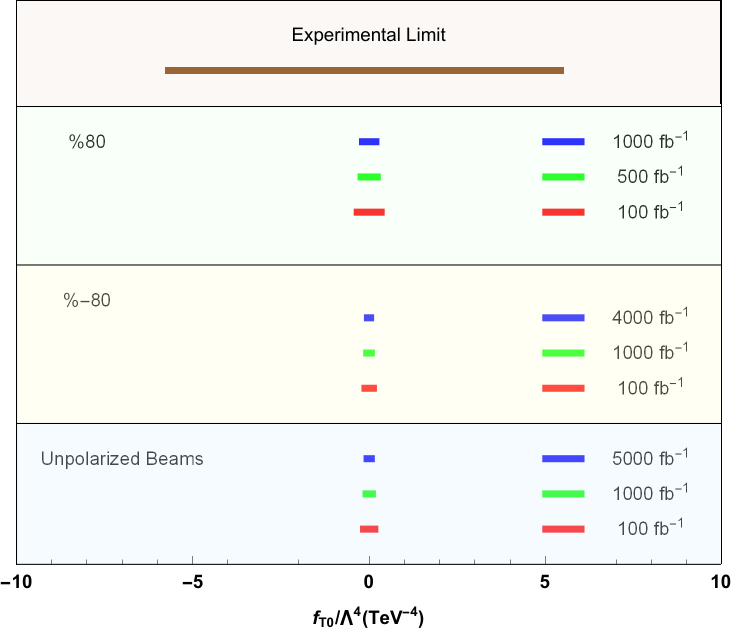}}}
\caption{ \label{fig:gamma} Comparison of experimental limits given in Ref.\cite{JHEP10-2021} and obtained sensitivity on $f_{T,0}/\Lambda^4$
for  the integrated luminosities of ${\cal L}=100, 500, 1000, 4000, 5000\hspace{0.8mm} fb^{-1}$. We take into account $P_{e^-}=-80\%, 0\%, 80\%$.
terms.}
\end{figure}

\begin{figure}[H]
\centerline{\scalebox{0.9}{\includegraphics{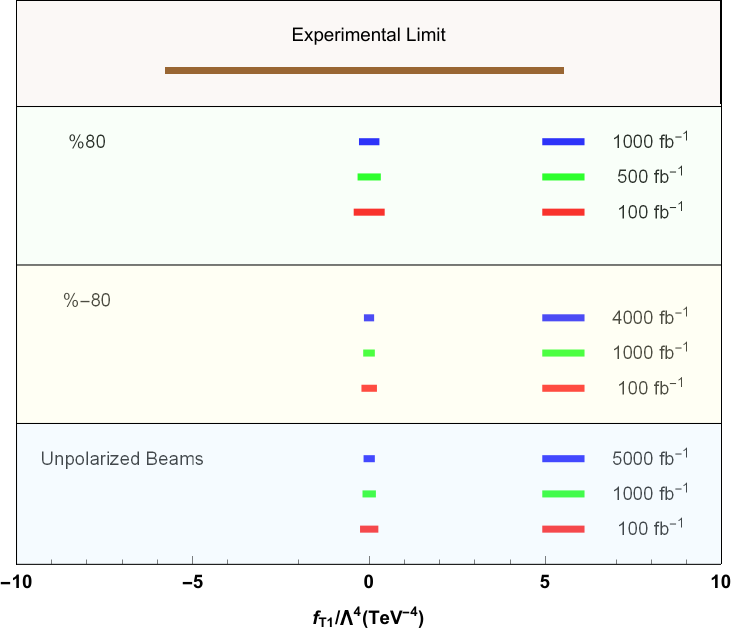}}}
\caption{ \label{fig:gamma} Same as in Fig. 6, but for $f_{T,1}/\Lambda^4$.}
\end{figure}

\begin{figure}[H]
\centerline{\scalebox{0.9}{\includegraphics{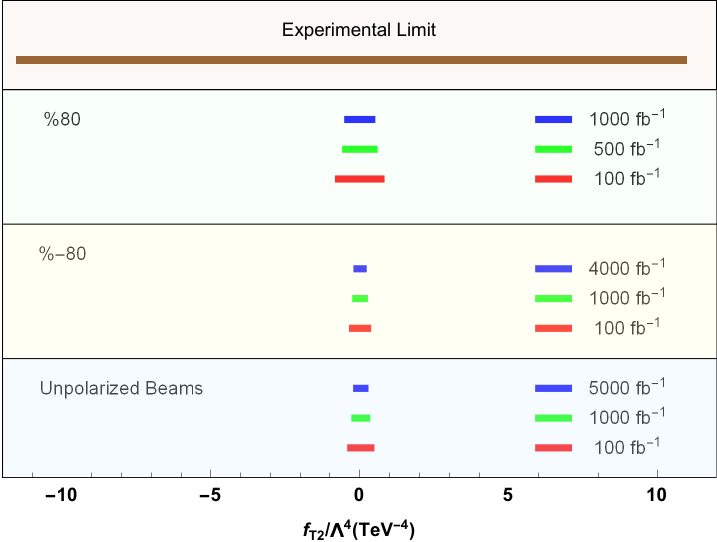}}}
\caption{ \label{fig:gamma} Same as in Fig. 6, but for $f_{T,2}/\Lambda^4$.}
\end{figure}

\begin{figure}[H]
\centerline{\scalebox{0.9}{\includegraphics{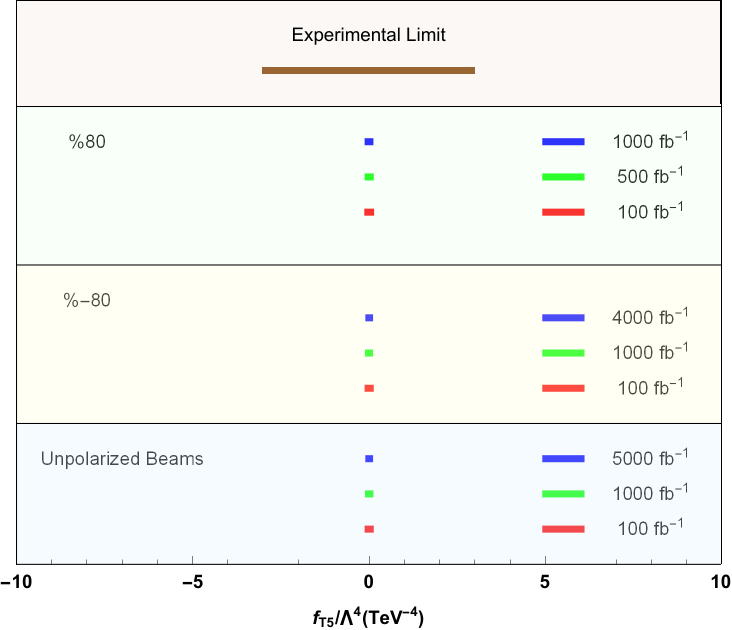}}}
\caption{ \label{fig:gamma}  Same as in Fig. 6, but for $f_{T,5}/\Lambda^4$.}
\end{figure}

\begin{figure}[H]
\centerline{\scalebox{0.9}{\includegraphics{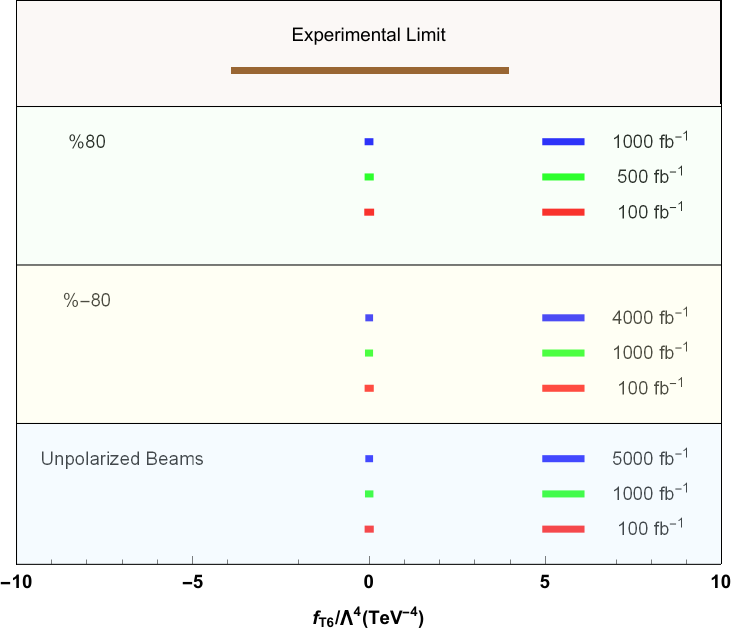}}}
\caption{ \label{fig:gamma} Same as in Fig. 6, but for $f_{T,6}/\Lambda^4$.}
\end{figure}

\begin{figure}[H]
\centerline{\scalebox{0.9}{\includegraphics{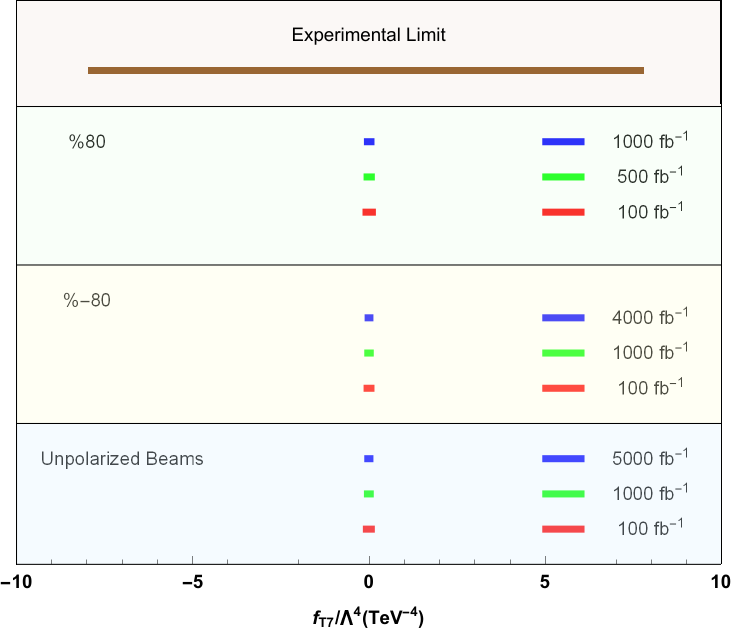}}}
\caption{ \label{fig:gamma} Same as in Fig. 6, but for $f_{T,7}/\Lambda^4$.}
\end{figure}

\begin{figure}[H]
\centerline{\scalebox{0.9}{\includegraphics{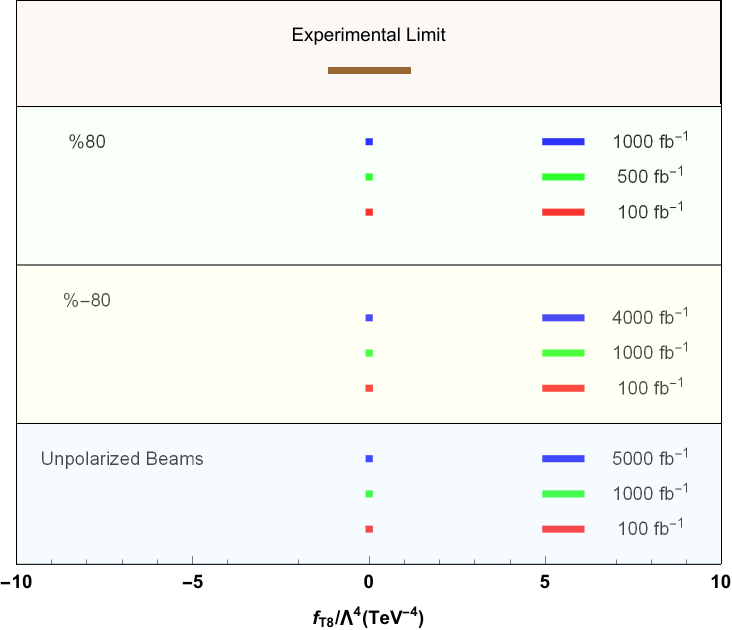}}}
\caption{ \label{fig:gamma} Same as in Fig. 6, but for $f_{T,8}/\Lambda^4$.}
\end{figure}

\begin{figure}[H]
\centerline{\scalebox{0.9}{\includegraphics{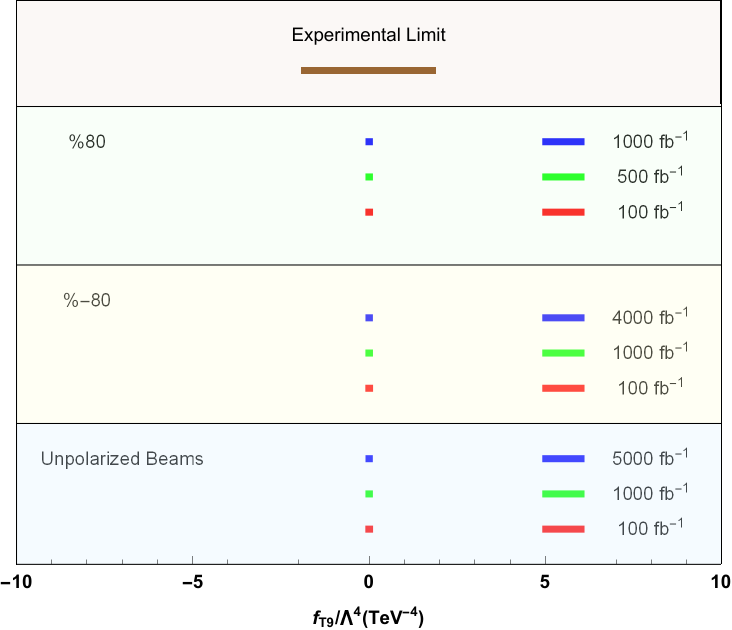}}}
\caption{ \label{fig:gamma} Same as in Fig. 6, but for $f_{T,9}/\Lambda^4$.}
\end{figure}

\end{document}